\documentclass[twocolumn,preprintnumbers,amsmath,amssymb]{revtex4}

\pdfoutput=1

\usepackage{todonotes}
\usepackage{SIunits,bm,amsmath}

\usepackage{graphicx}
\usepackage{epstopdf} 
\usepackage{dcolumn}
\usepackage{bm}
\usepackage{amsmath}

\usepackage{color} 

\renewcommand{\vec}[1]{\mathbf{#1}}

\newcommand{\ket}[1]
{
|#1 \rangle
}
\newcommand{\braket}[2]
{
\langle #1 | #2 \rangle
}

\newcommand{\obraket}[3]
{
\langle #1 | #2 | #3 \rangle
}

\begin{document}


\title{Comment on "Quantized Orbital Angular Momentum Transfer and Magnetic Dichroism in the Interaction of Electron Vortices with Matter"
}

\author{P. Schattschneider}
\affiliation{Institute of Solid State Physics, Vienna University of Technology, A-1040 Wien, Austria}

\affiliation{Ecole Centrale Paris, LMSSMat (CNRS UMR 8579), F-92295 Ch\^atenay-Malabry, France}
\email{schattschneider@ifp.tuwien.ac.at }

\author{S. L\"offler}
\affiliation{University Service Center for Transmission Electron Microscopy, Vienna University of Technology, A-1040 Wien, Austria}

\author{J. Verbeeck}
\affiliation{EMAT, University of Antwerp, Groenenborgerlaan 171, 2020 Antwerp, Belgium}

\begin{abstract}
Submitted to Physical Review Letters 11 July 2012. Accepted for publication 3 April 2013.
\end{abstract}


\maketitle
Lloyd {\it et al.}~\cite{LloydPRL2012} (LBY) find that - contrary to the case of optical vortices - transfer of orbital angular momentum (OAM) from vortex electrons to the  electronic degrees of freedom of an atom is possible, and thus explain the observed EMCD effect~\cite{VerbeeckNature2010, SchattPRB2012}. 

However, the experimental consequences discussed by LBY deserve a comment. We refer to equations in the said paper by "L", followed by the Eq. number; LBY's notation is adopted.

Concentrating on  electric dipole transitions in EELS we note that they do not depend on the condition
\begin{equation}
|\vec{q}| \ll |\vec{r} - \vec{R}|
\label{approx}
\end{equation}
 used to derive L8, see e.g.~\cite{Manson1972}. In fact, the electric dipole scattering kernel peaks at $|\vec{r}-\vec{R}| \sim |\vec{q}|$, i.e. when the probe electron passes close to  the  atom electron~\cite{SchattPRB99}. 
Therefore we use the exact  Hamiltonian L6.
In the experimentally relevant case of rigidly fixed atoms that LBY assume after L13, the atom is approximated by a spatial eigenstate at coordinate $\vec R_0$~\cite{Bethe}
$$
\braket{R}{\psi_n} \doteq \delta^3({\bf R}-{\bf R}_0)
$$
and we can  integrate  the nucleus contribution in Eq.~L7, leading to a transition matrix element between orthogonal initial and final internal states $\ket{\psi_e}$ and $\ket{\psi_e'}$
$$
\mathcal M_{if}=\obraket{\psi_e' \psi_B'}{H_{int}}{\psi_e \psi_B} .
$$
The vortex   $\ket{\psi_B}$ Eq.~L5 is a Bessel beam.
The  Hamiltonian depends on the electronic coordinate $\vec q$ in the center of mass system and the vortex coordinate $\vec r$ in the vortex centered system:
\begin{equation}
H_{int}=\frac{e^2}{4 \pi \varepsilon_0}\frac{1}{|{\bf r}-{\bf R}_0-{\bf q}|} \, .
\end{equation} 
The term in Eq.~L6 containing the atom coordinate ${\bf R}_0$ vanished because $\braket{\psi_e}{\psi_e'}=0$.
The matrix element L7 reads
\begin{eqnarray}
\mathcal M_{if}&=& \int d^2r \,  J_l(k_{\rho}r)J_{l'}(k_{\rho'}r)\int d^2q \, u(q) {u'}^\ast(q)  \nonumber \\
&& e^{i((m-m')\phi_q +(l-l')\Phi_r)}
\frac{1}{|{\bf r}-{\bf R}_0-{\bf q}|} .
\label{Mif}
\end{eqnarray}
$\phi_q , \, \Phi_r$ are the azimuthal angles of the respective vectors,  $u$ are the radial parts of the atom electron's wave function.  Substitution of  atom-centered coordinates ${\bf r}'={\bf r}-{\bf R}_0$ and the addition theorem for  Bessel functions $J_l$ yield
\begin{eqnarray}
\label{Mif2}
\mathcal M_{if}&=& \sum_{p,p',\rho, \rho'}J_{p}(k_{\rho}R_0)J_{p'}(k_{\rho'}R_0) \\
&&\int d^2r' \,  J_{l+p}(k_{\rho}r')J_{l'+p'}(k_{\rho'}r') \int d^2q \, u(q) {u'}^\ast(q)  \nonumber\\
&&  e^{i((m-m')\phi_q +(l+p-l'-p')\phi_r')}
\frac{1}{|{\bf r}'-{\bf q}|}. \nonumber 
\end{eqnarray}
Without loss of generality $\vec{R}_0$ points along the $x$ direction of the reference frame.
Both  azimuthal angles $\phi_q, \, \phi_r'$ refer to the  center of the atom.
Substituting $\varphi=\phi_r' - \phi_q$ in  the Coulomb term
$$
[r'^2+q^2-2 r' q \cos (\phi_r' - \phi_q)]^{-1/2} := F(r',q, (\phi_{r'} - \phi_q)) ,
$$
 LBY's azimuthal component Eq.~L10 reads
$$
{\mathcal M}_{az}= \int_0^{2 \pi} d\varphi e^{i \lambda \varphi} F(r',q,\varphi) \int_0^{2 \pi} d\phi_q e^{i(\lambda+\alpha)\phi_q}
$$
with   $\lambda=l+p-l'-p'$ and $\alpha=m-m'$.  
The second integral vanishes except for $\lambda=-\alpha$,
 giving rise to selection rules for dipole   transitions $\alpha= \pm 1$
\begin{equation}
l'=l \pm 1 +p-p' .
\label{Selection}
\end{equation}
Since $p-p'$ spans the  integer range,  dipole transitions  to any final vortex state $l'$  are possible; the outgoing probe electron is not in an OAM eigenstate~\footnote{This is a consequence of the uncertainty relation: The localisation of the atom electron to an angle $\Delta \Phi_R \sim |\vec q|/|\vec R_0|$ as seen from the vortex center induces an uncertainty $\Delta l' \geq 1/\Delta \Phi_R$, see~\cite{FrankeArnold2004}.}.

The conclusion of LBY must therefore be modified: electric dipole transitions mediate the transfer of OAM, but in general, the transfer is not quantized.

To reconcile this result with the findings of LBY and  with the experimental evidence~\cite{VerbeeckNature2010} that vortices can be used to detect chiral electronic transitions it is sufficient to reconsider  Eq.~\ref{Mif2} in the limit $k_{\rho} R_0 \rightarrow 0$: Since $J_p(0)$ vanishes except for $p=0$ the sum over $p, \, p'$ collapses into a single term,  and the selection rule Eq.~\ref{Selection} reads
$
l'=l -\alpha  
\label{Selection2}
$
which is  L12  for $L=L'$.
The matrix elements violating these selection rules will be small for small displacements $|\vec R_0|$ of the atom from the vortex core. This means that the larger the observed cluster the fainter is the  EMCD effect. 

Furthermore, an electron vortex  can exchange OAM with the crystal lattice so that neither the assumption of an incident  nor that of an outgoing OAM eigenstate are fulfilled in practice~\cite{LoefflerACA2012}.


{\bf Acknowledgements: }

P.S. and S.L. acknowledge financial support by the Austrian Science Fund, project I543-N20. 
J.V. acknowledges funding from the European Research Council under the
7th Framework Program (FP7), ERC grant N246791 - COUNTATOMS and ERC
Starting Grant 278510 VORTEX.

%

\end{document}